\documentstyle[11pt,epsf]{article}

\def\draftversion{N}

\if \draftversion Y

\fi

\newcommand{\reseteqnum}{\setcounter{equation}{0}}

\newcommand{\plb}[3]{Phys. Lett. {\bf B#1} (#2) #3} 
\newcommand{\prl}[3]{Phys. Rev. Lett. {\bf #1} (#2) #3}
\newcommand{\prd}[3]{Phys. Rev. {\bf D#1} (#2) #3}
\newcommand{\npb}[3]{Nucl. Phys. {\bf B#1} (#2) #3}

\newcommand{\sla}[1]{{\ooalign{\hfil/\hfil\crcr$#1$}}}
\newcommand{\f}[2]{\frac{#1}{#2}}

\newcommand{\lb}[0]{\l[}  
\newcommand{\rb}[0]{\r]}
\newcommand{\lc}[0]{\l\{} 
\newcommand{\rc}[0]{\r\}}

\newcommand{\be}{\begin{eqnarray}}
\newcommand{\ee}{\end{eqnarray}}
\newcommand{\bc}{\begin{center}}
\newcommand{\ec}{\end{center}}
\newcommand{\vi}[2]{\mbox{$V_{1{#1}}\left({#2}\right)$}}
\newcommand{\vii}[2]{\mbox{$V_{2{#1}}\left({#2}\right)$}}
\newcommand{\vid}[2]{\mbox{$V^\dagger_{1{#1}}\left({#2}\right)$}}
\newcommand{\viid}[2]{\mbox{$V^\dagger_{2{#1}}\left({#2}\right)$}}

\def\l{\left}
\def\r{\right}

\def\kph{\displaystyle{k+\f{p}{2}}}
\def\kphm{\displaystyle{k-\f{p}{2}}}

\def\intk{\int_{-\f{\pi}{a}}^{\f{\pi}{a}}\f{d^4k}{(2\pi)^4}}

\title{
One-loop Analyses of lattice QCD with the overlap Dirac operator
\author{
Masato Ishibashi\thanks{e-mail address:
ishibash@hep-th.phys.s.u-tokyo.ac.jp}\
, \ Yoshio Kikukawa\thanks{e-mail address:
kikukawa@eken.phys.nagoya-u.ac.jp} \ 
, \ Tatsuya Noguchi\thanks{e-mail address:
noguchi@gauge.scphys.kyoto-u.ac.jp} \\
and \ Atsushi Yamada\thanks{e-mail address:
atsushi@hep-th.phys.s.u-tokyo.ac.jp} 
\\
\\
{\normalsize\em * {\S} Department of Physics,University of Tokyo}\\
{\normalsize\em Tokyo 113-0033, Japan}
\\
{\normalsize\em {\dag} Department of Physics, Nagoya University 
}\\
{\normalsize\em Nagoya 464-8602, Japan}
\\
{\normalsize\em {\ddag} Department of Physics, Kyoto University 
}\\
{\normalsize\em Kyoto 606-8502, Japan}
\\
\\
\date{\normalsize November, 1999}
}
}

\begin{document}
\maketitle

\begin{abstract}
We discuss the weak coupling expansion of lattice QCD with the overlap
 Dirac operator.
 The Feynman rules for lattice QCD with the overlap 
Dirac operator are derived and the quark self-energy and vacuum
 polarization are studied at the one-loop level. We confirm that
their divergent parts agree with those in the continuum theory. 
\end{abstract}

\begin{quote}
\end{quote}

\newpage
\section{Introduction} 
\label{sec:introduction}
\reseteqnum

The overlap Dirac operator~\cite{overlap-D} satisfies the
Ginsparg-Wilson relation~\cite{gw} and thus keeps the lattice chiral
symmetry proposed in Ref. \cite{luscher}.
This Dirac operator has no species doublers and is local (with
exponentially decaying tails) for gauge fields with small field strength
\cite{locality-of-overlap-D}.
Originally the overlap Dirac operator is derived from the 
overlap formalism \cite{overlap,odd-dim-overlap}.
In this formalism the axial anomaly is calculated 
and the renormalizability is discussed within the perturbation 
theory \cite{daemi-strathdee,yamada}. 
General consequences of the analyses performed in the overlap 
formalism are expected to hold also in the case of the overlap 
Dirac operator. However the actual procedures of the perturbative 
analyses are so different in these two cases and 
the correspondence between them is not straightforward. 
In the overlap formalism, time independent perturbation 
theory using creation and annihilation operators should be invoked,
and the perturbation theory is not conveniently expressed as Feynmann
rules, while such diagrammatic techniques are available for the 
overlap Dirac operator. Moreover, the finite parts of renormalization
factors on the lattice, which are necessary to extract physical
observables from numerical simulations, are 
different \cite{weak-coupling-exp}. 

Therefore in this paper, we discuss the weak coupling expansion of
lattice QCD using the overlap Dirac operator. Our study is an
extension of the previous analyses of Ref. \cite{kikukawa-yamada},
where only the axial anomaly is computed.   
\footnote{
The axial anomaly has been 
discussed using a different calculational scheme in 
Ref.~\cite{fujikawa,adams,suzuki}.}
We derive Feynman rules for lattice QCD with the overlap Dirac operator
and compute the quark self-energy and vacuum polarization. 
We demonstrate that non-local divergent terms, which are not
necessarily forbidden by the Ginsparg-Wilson relation, are not
in fact induced in the quark self-energy. 
Once non-local divergent terms are known to
be absent, the Ginsparg-Wilson relation prohibits divergent mass terms,
as is shown later. 
The divergent parts of the wave function 
renormalization factors agree with the continuum
theory. We also consider the vacuum polarization. Again, non-local
divergences are absent, and the wave function renormalization factors
are correctly reproduced.    
 
Our analyses are consistent with Ref.\cite{renormalizability-GW-QCD},
where a formal argument on the proof of the renormalizability 
was given for a general Dirac operator satisfying the Ginsparg-Wilson 
relation.
Supplementing their formal argument,
we take the overlap Dirac operator and show explicitly the nontrivial 
cancellations 
necessary to ensure renormalizability at one-loop level.
In Ref.\cite{lambda-parameter-overlapQCD} 
the finite part of the vacuum polarization 
at one-loop has been evaluated 
and the lambda parameter ratio has been obtained. But the quark
self-energy, which determines the chiral property 
of the renormalized fermions, was not analyzed.

\section{Derivation of the Feynman rules}
\label{sec:derivation-of-feynman-rule}
\reseteqnum
In this section we derive the Feynman rules for lattice QCD 
with the overlap Dirac operator, which are necessary for 
the one-loop analysis. 
The fermion action with the overlap Dirac operator is   
\begin{eqnarray}
S_F &=& a^4\sum_m \bar{\psi}(m) \, D \, \psi(m),
\end{eqnarray}
where
\begin{eqnarray}
\label{odo}
D &=& \f{1}{a}\l( 1 + X\f{1}{\sqrt{X^\dagger X}}\r).
\end{eqnarray}
Here $X$ is the Wilson-Dirac operator defined as
\begin{eqnarray}
X_{mn}&=&\f{1}{2a}\sum_{\mu =1}^{4}\lb \gamma_\mu\lc\delta_{m+\hat \mu, n}
U_\mu(m) - \delta_{m, n+\hat\mu}U_\mu^\dagger(n)\rc \r.\nonumber\\
&&\l. + r\lc2\delta_{m,n}-\delta_{m+\mu, n}
U_\mu(m) - \delta_{m, n+\mu}U_\mu^\dagger(n)\rc\rb + \f{M_0}{a}
\delta_{m,n},\nonumber\\
\end{eqnarray} 
where $U_\mu = e^{iagA_\mu}$ is the link variable 
and $r$ is the Wilson parameter. 
Expanding $X_{mn}$ up to the second order in the coupling 
constant $g$, we obtain

\begin{eqnarray}
  X_{mn} &=& \int^{\f{\pi}{a}}_{-\f{\pi}{a}}\f{d^4p}{(2\pi)^4}\f{d^4q}{(2\pi)^4}e^{ia(qm -
  pn)}X(q,p),\\
  X(q,p) &=& X_0(p)(2\pi)^4 \delta_P(q - p) + X_1(q,p) + X_2(q,p) +
  {\cal O}(g^3),
\end{eqnarray}
where 
\begin{eqnarray}
  X_0(p) &=& \sum_{\rho}\f{i}{a}\gamma_\rho\sin ap_\rho + 
\f{r}{a}\sum_\mu(1- \cos a
  p_\mu) + \f{1}{a}M_0,\label{eq:X_0}\\ 
X_1(q,p) &=& \sum_{A,\mu}\intk (2\pi)^4\delta_P(q-p-k)\nonumber\\
&&\quad\times g A_\mu^A(k)T^A V_{1\mu}\l(p+\f{k}{2}\r),\label{eq:X_1}\\ 
X_2(q,p) &=& \sum_{A,B,\mu,\nu}\int^{\f{\pi}{a}}_{-\f{\pi}{a}}
\f{d^4k_1}{(2\pi)^4}
\f{d^4k_2}{(2\pi)^4}
(2\pi)^4\delta_P(q-p-\sum  k_i)\nonumber\\ 
&&\quad\times\f{g^2}{2}A_\mu^A(k_1)A_\mu^B(k_2)T^AT^BV_{2\mu}\l(p+\f{\sum k_i}{2}\r), \label{eq:X_2}
\end{eqnarray}
and the index $i$ runs 1,2 and $\delta_P$ is the periodic lattice 
delta function. $T^A$ are $SU(3)$ generators in the fundamental 
representation.  
The vertex functions $V_{1\mu}$ and $V_{2\mu}$ are given by 
\begin{eqnarray}
V_{1\mu}(p + \f{k}{2}) &=& i\gamma_\mu\cos a\l(p +
\f{k}{2}\r)_\mu + r\sin a\l(p  + \f{k}{2}\r)_\mu,\label{eq:V_1}\nonumber\\
V_{2\mu}\l(p + \f{\sum k_i}{2}\r) &=& -i\gamma_\mu a \sin a\l(p +
\f{\sum k_i}{2}\r)_\mu + ar\cos a\l(p + \f{\sum k_i}{2}\r)_\mu.
\label{eq:V_2}\nonumber\\
\end{eqnarray}

The weak coupling expansion of the overlap Dirac operator is derived  
 in the following way \cite{kikukawa-yamada}.
Using the following identity:
  
\begin{eqnarray}
  \f{1}{\sqrt{X^\dagger X}} = \int^\infty_{-\infty}\f{dt}{\pi}\f{1}{t^2 +
  X^\dagger X},
\end{eqnarray}
we expand the r.h.s of Eq.~(\ref{odo}) up to the second order in $g$ as
\begin{eqnarray}
\label{eq:exp}  
aD &=& aD_0\nonumber\\
&& + \int^\infty_{-\infty}\f{dt}{\pi}\f{1}{t^2 + X_0^\dagger X_0}
\l(t^2 X_1 - X_0 X^\dagger_1 X_0\r)\f{1}{t^2 + X_0^\dagger X_0}\nonumber\\
&& + \int^\infty_{-\infty}\f{dt}{\pi}\f{1}{t^2 +
  X_0^\dagger X_0}\l(t^2 X_2 - X_0 X^\dagger_2 X_0\r)\f{1}{t^2 +
  X_0^\dagger X_0}\nonumber\\
&& - \int^\infty_{-\infty}\f{dt}{\pi}t^2\f{1}{t^2 +
  X_0^\dagger X_0}(X_1)\f{1}{t^2 + X_0^\dagger X_0}\l(X^\dagger_0 X_1
+ X^\dagger_1 X_0\r)\f{1}{t^2 +X_0^\dagger X_0}\nonumber\\
&& -\int^\infty_{-\infty}\f{dt}{\pi}t^2\f{1}{t^2 +
  X_0^\dagger X_0}\l(X_0 X^\dagger_1\r)\f{1}{t^2 + X_0^\dagger X_0}(X_1)\f{1}{t^2 + X_0^\dagger X_0}\nonumber\\
&& +\int^\infty_{-\infty}\f{dt}{\pi}(X_0)\f{1}{t^2 +
  X_0^\dagger X_0}\l(X_1^\dagger X_0\r)\f{1}{t^2 + X_0^\dagger
  X_0}\l(X_1^\dagger X_0\r)\f{1}{t^2 + X_0^\dagger X_0} + \cdots,\nonumber\\
\end{eqnarray}
where $D_0$ is the zeroth-order term in $g$. The inverse of $D_0$ 
in momentum space is
\begin{eqnarray}
  D_0^{-1}(p) &=& \f{-i\gamma_\rho\sin a
  p_\rho}{2(\omega(p)+b(p))} + \f{a}{2} = \f{a}{2}\f{X^\dagger_0(p) + \omega(p)}{\omega(p) + b(p)},\label{eq:D_0^{-1}}\\
b(p) &=& \f{r}{a}\sum_\mu(1 - \cos a p_\mu) + \f{1}{a}M_0,\\
  a\omega(p) &=& \sqrt{\sin^2 ap_\mu + \l(r\sum_\mu(1-\cos a p_\mu) + M_0\r)^2}
>0.\label{eq:w}
\end{eqnarray}
For $-2r < M_0 < 0$ the propagator $D_0^{-1}(p)$ exibits a masssless 
pole only when 
$p_\mu =0$, and there are no doublers. In the following, we take
$M_0$ in this region. We also note that
the quark fields $\psi$ are not properly normalized since 
the continuum limit of $D_0$ is $i{\ooalign{\hfil/\hfil\crcr$p$}}
/|M_0|$.

\begin{figure}[htdp]
\begin{center}
\leavevmode
\epsfxsize=40mm
\epsfbox{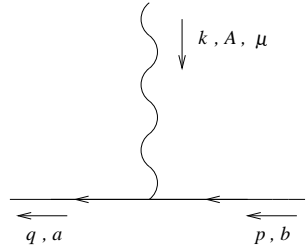}
\caption{Gluon-fermion theree-point vertex}
\label{fig:3pt}
\end{center}
\end{figure}

\begin{center}
\begin{figure}[htdp]
\leavevmode
\epsfxsize=59mm
\epsfbox{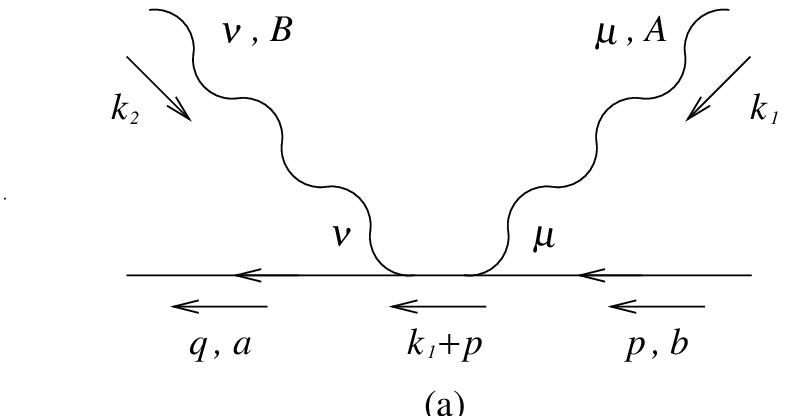}
\epsfxsize=59mm
\epsfbox{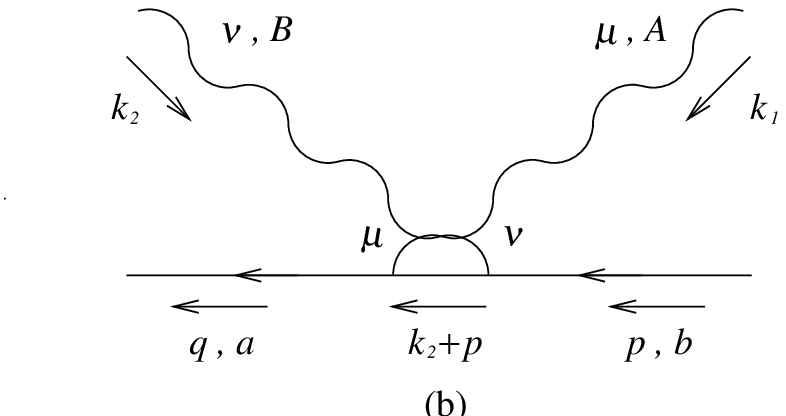}
\caption{Gluon-fermion four-point vertices (a) and (b)}
\label{fig:4pt1}
\end{figure}
\end{center}

After performing Fourier transformation in Eq.(\ref{eq:exp}), we obtain the overlap Dirac
operator in momentum space
up to the second order in $g$:
\begin{eqnarray}
  D(q,p) &=& D_0(p)(2\pi)^4\delta_P(q-p) + V(q,p),
\end{eqnarray}
where 
\begin{eqnarray}
V(q,p)
&=& \lc \f{1}{\omega(q) + \omega(p)}\rc\lb X_1(q,p) - \f{X_0(q)}{\omega(q)}X^\dagger_1(q,p)\f{X_0(p)}{\omega(p)}\rb\nonumber\\
&& + \lc \f{1}{\omega(q) + \omega(p)}\rc\lb X_2(q,p) - \f{X_0(q)}{\omega(q)}X^\dagger_2(q,p)\f{X_0(p)}{\omega(p)}\rb\nonumber\\
&& + \int_{r}
\lc \f{1}{\omega(q) + \omega(p)}\rc
\lc \f{1}{\omega(q) + \omega(r)}\rc
\lc \f{1}{\omega(r) + \omega(p)}\rc\nonumber\\
&&\times\Biggl[- X_0(q)X^\dagger_1(q,r)X_1(r,p)\nonumber\\
&& \quad -X_1(q,r)X_0^\dagger(r)X_1(r,p) - X_1(q,r)X^\dagger_1(r,p)X_0(p)\nonumber\\
&& \quad +\f{\omega(q) + \omega(p) +\omega(r)}{\omega(q)\omega(p)\omega(r)}X_0(q)X^\dagger_1(q,r)X_0(r)X^\dagger_1(r,p)X_0(p)\Biggr]
 +\cdots.\nonumber\\
\end{eqnarray}
From this weak coupling expansion we can derive
the Feynman rules for lattice QCD with the overlap Dirac operator.
For the quark propagator we assign the expression (\ref{eq:D_0^{-1}}). 
For the quark-gluon three-point vertex depicted in Fig. \ref{fig:3pt},
we assign
\begin{eqnarray}
 -\f{g}{a(\omega(q) +\omega(p))}T^A_{ab}\l\{ 
V_{1\mu}\l(p + \f{k}{2}\r) - \f{X_0(q)}{\omega(q)}V_{1\mu}^\dagger\l(p +
  \f{k}{2}\r)\f{X_0(p)}{\omega(p)}\r\}.
\end{eqnarray}
For the quark-gluon four-point vertex
depicted in Fig.(\ref{fig:4pt1}), we assign 

\begin{eqnarray}
 &&  \f{-g^2}{2a}\l\{T^A,T^B\r\}_{ab} \lb\f{1}{\omega(q) + \omega(p)}\lc V_{2\mu}\l(p
  + \f{k}{2}\r)\delta_{\mu\nu} -
  \f{X_0(q)}{\omega(q)}V_{2\mu}^\dagger\l(p +
  \f{k}{2}\r)\f{X_0(p)}{\omega(p)} \delta_{\mu\nu}\rc\r.\nonumber\\ 
 &&\;\;\;-\f{1}{\{\omega(q) + \omega(p)\}\{\omega(p) + \omega(p+k_1)\}\{\omega(p+k_1) + \omega(q)\}}\nonumber\\
 && \times \l\{ X_0(q)V^\dagger_{1\mu}\l(p+k_2 +
  \f{k_1}{2}\r)V_{1\nu}\l(p + \f{k_2}{2}\r) + V_{1\mu}\l(p+k_2 +
  \f{k_1}{2}\r)X_0^\dagger(p+k_2) V_{1\nu}(p + \f{k_2}{2})\r.\nonumber\\
 &&  \hspace{0.5cm}+  V_{1\mu}\l(p+k_2 + \f{k_1}{2}\r)
  V_{1\nu}^\dagger\l(p + \f{k_2}{2}\r)X_0(p)\nonumber \\
&&  \hspace{0.5cm}- \l.\f{\omega(q) +\omega(p) +\omega(p+k_2)}{\omega(q)\omega(p)\omega(p+k_2)}X_0(q)V_{1\mu}^\dagger\l(p+k_2 + \f{k_1}{2}\r)X_0(p+k_2)V_{1\nu}^\dagger\l(p + \f{k_2}{2}\r)X_0(p)\r\}\nonumber \\
 &&\;\;\;-\f{1}{\{\omega(q) + \omega(p)\}\{\omega(p) + \omega(p+k_1)\}\{\omega(p+k_1) + \omega(q)\}}\nonumber \\
 && \times \l\{ X_0(q)V^\dagger_{1\nu}\l(p+k_1 +
  \f{k_2}{2}\r)V_{1\mu}\l(p + \f{k_1}{2}\r) + V_{1\nu}\l(p+k_1 +
  \f{k_2}{2}\r)X_0^\dagger(p+k_1) V_{1\mu}(p + \f{k_1}{2})\r.\nonumber \\
 &&  \hspace{0.5cm}+  V_{1\nu}\l(p+k_1 + \f{k_2}{2}\r) V_{1\mu}^\dagger\l(p + \f{k_1}{2}\r)X_0(p)\nonumber \\
&&  \hspace{0.5cm}- \l.\l.\f{\omega(q) +\omega(p) +\omega(p+k_1)}{\omega(q)\omega(p)\omega(p+k_1)}X_0(q)V_{1\nu}^\dagger\l(p+k_1 + \f{k_2}{2}\r)X_0(p+k_1)V_{1\mu}^\dagger\l(p + \f{k_1}{2}\r)X_0(p)\r\}\rb.\nonumber\\
\end{eqnarray}    

The gauge field action is chosen as Wilson's plaquette 
action, where 
\begin{equation}
  S_G = \frac{1}{g^2}\sum_m\sum_{\mu\nu} 
{\rm Tr} \left(1- U_{\mu\nu}(m) \right),
\end{equation}
where $U_{\mu\nu}(m)=U_\mu(m)U_\nu(m+\hat \mu) 
U_\mu(m+\hat \nu)^\dagger U_\nu(m)^\dagger $.\footnote{
L\"uscher proposed other action for gauge field in the abelian gauge 
theory\cite{abelian}, which seems to be more appropriate in view
of the locality properties.}
The gluon propagator is given by 

\begin{eqnarray}
G_{\mu\nu}(k) &=& \f{1}{\tilde k^2}\left( \delta_{\mu\nu} - (1 - \alpha)
\f{\tilde k_\mu \tilde k_\nu}{\tilde k^2}\right),
\end{eqnarray}
where $\tilde k_\mu =(2/a)\sin ak_\mu/2$, and $\alpha$ is the
gauge fixing parameter .

In the following sections we calculate the quark self-energy and the
vacuum polarization at one-loop level using these Feynman rules.

\section{Evaluation of the quark self-energy}
\label{sec:fermion-self-energy}
\reseteqnum
In this section we evaluate the one-loop quark self-energy of the lattice QCD
with the overlap Dirac operator and we show that the structure of divergence 
is equal to that in the continuum theory.   
In the lattice QCD with the overlap Dirac operator,
there exists the chiral symmetry based on the Ginsparg-Wilson relation as
$\{\gamma_5, D\}=aD\gamma_5D$.
First, let us investigate what the Ginsparg-Wilson relation tells us about 
the structure of divergence. Up to one-loop order the Ginsparg-Wilson relation
is written as     
 
\begin{eqnarray}
  \lc\gamma_5 , S_F^{-1}(p) + \Sigma(p)\rc = a( S_F^{-1}(p) +
  \Sigma(p))\gamma_5( S_F^{-1}(p) + \Sigma(p)),
\end{eqnarray}
where $S_F(p)$ is the quark propagator and $\Sigma(p)$ is
the one-loop quark self-energy. Ignoring the second-order term and using
the Ginsparg-Wilson relation at the tree level, we obtain 
\begin{eqnarray}
\label{eq1} 
\lc\gamma_5,\Sigma(p)\rc &=& a \Sigma(p)\gamma_5 
  S_F^{-1}(p) + a S_F^{-1}(p)\gamma_5 \Sigma(p).
\end{eqnarray}
Since $\Sigma(p)$ has dimension 1, the possible structure of divergence
is
\begin{eqnarray}
\label{eq2}
\Sigma(p)&=&\sum_{n=1}^\infty\lc\f{i\sla{p}C_{1n}}{(a^2p^2)^n} +
\f{ap^2C_{2n}}{(a^2p^2)^{n+1}}\rc + \f{C_3}{a} + \sum_{n=1}^\infty
C_{4n}i\sla{p}(\log{a^2p^2})^n \nonumber\\
&&+ finite\; terms,
\end{eqnarray}
where $C_{1n}, C_{2n}$ ,$C_{3}$ and $C_{4n}$ is generally
a linear combination of constants and the power of $\log{a^2p^2}$.
In the above equation the first and second terms are non-local
divergences and the third and fourth terms are local divergences.
In general, the Ginsparg-Wilson relation does not exclude non-local
divergent terms, which spoil the renormalizability of the theory.  
Now we compute the quark self-energy $\Sigma$ with the overlap Dirac operator
and we show that non-local divergent terms don't appear. If such 
non-local terms are absent, the linearly divergent terms $C_3$ in Eq.
(\ref{eq2}) should also vanish due to the Ginsparg-Wilson relation,
which we elucidate in our calculation.
We also compute
the logarithmic divergent part and show that the divergent part of 
the quark field renormalization factor
is equal to that in the continuum theory.

\begin{figure}[htdp]
\begin{center}
\leavevmode
\epsfxsize=50mm
\epsfbox{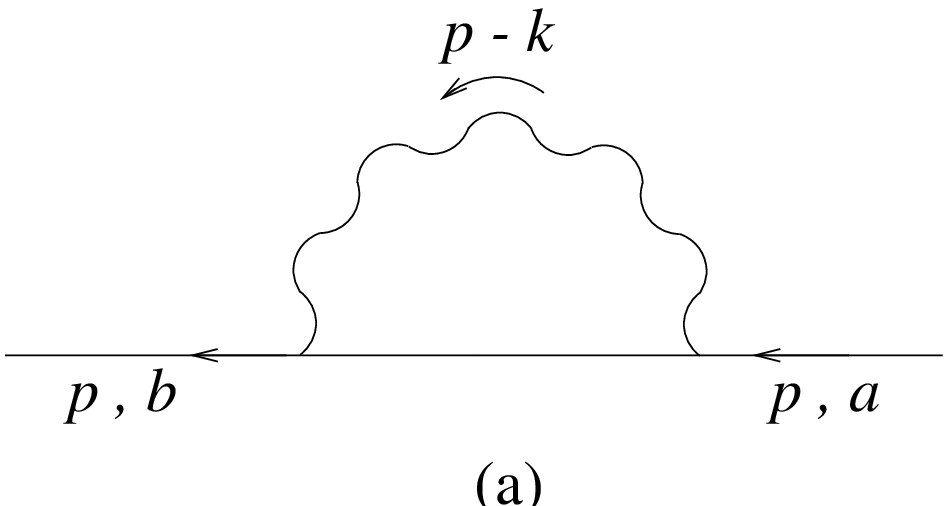}
\epsfxsize=50mm
\hspace{1.5cm}
\epsfbox{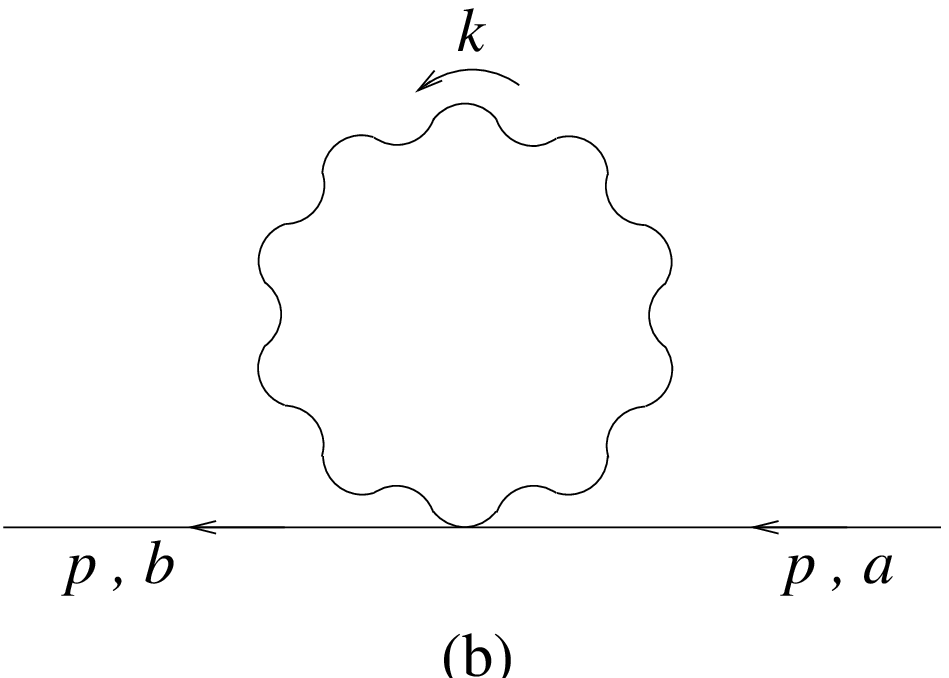}
\caption{Feynman diagrams for the quark self-energy}
\label{fig:fermion-se1}
\end{center}
\end{figure}

The Feynman diagrams which contribute to the one-loop quark self-energy
 are shown in Fig.~\ref{fig:fermion-se1}.
The amplitude corresponding to Fig.3(a) is
\begin{eqnarray} 
\Sigma^{(a)}(p)&= &   \f{g^2}{a^2}\sum_{A}(T^{A})^{2}_{ba}\sum_{\mu\nu}
\intk G_{\mu\nu}(p-k)
\nonumber
\\
&&\times\f{1}{\lc\omega(k)
  +\omega(p)\rc^2}\lc\vi{\mu}{\displaystyle{\f{p+k}{2}}} - 
\f{X_0(p)}{\omega(p)}\vid{\mu}{\displaystyle{\f{p+k}{2}}}\f{X_0(k)}{\omega(k)}
\rc\nonumber\\
&&\times \lc\f{-i\gamma_{\rho}\sin{ak_{\rho}}}{2(\omega(k)+b(k))}
+\f{a}{2}\rc \lc\vi{\nu}{\displaystyle{\f{p+k}{2}}} -
\f{X_0(k)}{\omega(k)}\vid{\nu}{\displaystyle{\f{p+k}{2}}}\f{X_0(p)}
{\omega(p)}\rc.\nonumber\\
\end{eqnarray}
The amplitude corresponding to Fig.3(b) is
\begin{eqnarray} 
\Sigma^{(b)}(p)&=& -  \f{g^2}{2a^2}\sum_{A}(T^{A})^{2}_{ba}\sum_{\mu\nu}
\intk G_{\mu\nu}(k) 
\nonumber
\\
&&\lb\f{1}{2\omega(p)}\lc V_{2\mu}(p) -
  \f{X_0(p)}{\omega(p)}V_{2\mu}^\dagger(p)\f{X_0(p)}{\omega(p)}\rc\delta_{\mu\nu}\r. 
 -\f{1}{\omega(p)\lc\omega(p) + \omega(p+k)\rc^2}\nonumber \\
 && \times \l\{ X_0(p)V^\dagger_{1\mu}\l(p+
  \f{k}{2}\r)V_{1\nu}\l(p + \f{k}{2}\r) + V_{1\mu}\l(p+
  \f{k}{2}\r)X_0^\dagger(p+k) V_{1\nu}\l(p + \f{k}{2}\r)\r.\nonumber \\
 &&  \hspace{0.5cm}+  V_{1\mu}\l(p+\f{k}{2}\r) V_{1\nu}^\dagger\l(p + \f{k}{2}\r)X_0(p)\nonumber \\
&&  \hspace{0.5cm}-
  \l.\l.\f{2\omega(p)+\omega(p+k)}{\omega(p)^2\omega(p+k)}X_0(p)V_{1\mu}^\dagger\l(p+\f{k}{2}\r)X_0(p+k)V_{1\nu}^\dagger\l(p + \f{k}{2}\r)X_0(p)\r\}\rb\nonumber.\\
\end{eqnarray}
To evaluate the divergent parts of $\Sigma^{(a)}$ and $\Sigma^{(b)}$,
we rescale the integration momentums $k_{\mu}\to k_{\mu}/a$ 
in each amplitude and we 
obtain
 
\begin{eqnarray} 
\Sigma^{(a)}(p)&= &   \f{\bar{g}^2}{2a}\sum_{\mu\nu}\int_{k}G_{\mu\nu}(pa-k)
\nonumber
\\
&&\times\f{1}{\lc\omega(k)
  +\omega(pa)\rc^2}\lc\vi{\mu}{\displaystyle{\f{pa+k}{2}}} - 
\f{X_0(pa)}{\omega(pa)}\vid{\mu}{\displaystyle{\f{pa+k}{2}}}\f{X_0(k)}
{\omega(k)}
\rc\nonumber\\
&&\times \lc\f{-i\gamma_{\rho}\sin{k_{\rho}}}{(\omega(k)+b(k))}
+1\rc \lc\vi{\nu}{\displaystyle{\f{pa+k}{2}}} -
\f{X_0(k)}{\omega(k)}\vid{\nu}{\displaystyle{\f{pa+k}{2}}}\f{X_0(pa)}
{\omega(pa)}\rc,\label{eq:sigma(a)}\nonumber\\
\end{eqnarray}
and
\begin{eqnarray} 
\Sigma^{(b)}(p)&= & -  \f{\bar{g}^2}{2a}\sum_{\mu\nu}\int_{k}G_{\mu\nu}(k)
\f{1}{\omega(pa)}
\nonumber
\\
&&\lb\f{1}{2}\lc V_{2\mu}(pa) -
  \f{X_0(pa)}{\omega(pa)}V_{2\mu}^\dagger(pa)\f{X_0(pa)}{\omega(pa)}\rc\delta_{\mu\nu}\r. 
 -\f{1}{\lc\omega(pa) + \omega(pa+k)\rc^2}\nonumber \\
 && \times \l\{ X_0(pa)V^\dagger_{1\mu}\l(pa+
  \f{k}{2}\r)V_{1\nu}\l(pa + \f{k}{2}\r) + V_{1\mu}\l(pa+
  \f{k}{2}\r)X_0^\dagger(pa+k) V_{1\nu}\l(pa + \f{k}{2}\r)\r.\nonumber \\
 &&  \hspace{0.5cm}+  V_{1\mu}\l(pa+\f{k}{2}\r) V_{1\nu}^\dagger\l(pa + \f{k}{2}\r)X_0(pa)\nonumber \\
&&  \hspace{0.5cm}-
  \l.\l.\f{2\omega(pa)+\omega(pa+k)}{\omega(pa)^2\omega(pa+k)}X_0(p)V_{1\mu}^\dagger\l(pa+\f{k}{2}\r)X_0(pa+k)V_{1\nu}^\dagger\l(pa + \f{k}{2}\r)X_0(pa)\r\}\rb,\nonumber\label{eq:sigma(b)}\\
\end{eqnarray}
where $\bar{g}^2\delta_{ab}\equiv g^2\sum_{A}(T^{A})^{2}_{ba}$ \footnote
{When $T^A$ are the $SU(N)$ generators in the fundamental
representation, 
$\bar{g^2}$ is $\f{N^2-1}
{2N}g^2$.} and 
$\int_{k}\equiv \int^{\pi}_{-\pi}
d^4k /(2\pi)^4$. In the above two equations $\omega$, $X_0$, $V_{1\mu}$,
$V_{2\mu}$ and $G_{\mu\nu}$ are appropriately redefined according to
the rescaling of $k_{\mu}$.
For example 
$\omega(k) = \sqrt{\sin^2 k_\mu + \l(r\sum_\mu(1-\cos k_\mu) +
M_0\r)^2}$.
First, we show that there are not the nonlocal divergences of
the forms 
$i\sla{p}C_{1n}/(a^2p^2)^n$ and $ap^2C_{2n}/(a^2p^2)^{n+1}$. 
For this purpose we
see that Eq.(\ref{eq:sigma(a)}) and (\ref{eq:sigma(b)})
are not singular for $p=0$. Setting $p=0$ in them, we have 
\begin{eqnarray} 
\Sigma^{(i)}(0)&=& \f{\bar{g}^2}{2a}\sum_{\mu\nu}\int_{k}G_{\mu\nu}(k)
\f{1}{\lc\omega(k)-M_0\rc^2}\sigma_{\mu\nu}^{(i)}(k),\quad i = a ,b.
\label{eq3}
\end{eqnarray}
where
\begin{eqnarray}
\sigma_{\mu\nu}^{(a)}(k) &=&\lc\vi{\mu}{\displaystyle{\f{k}{2}}} + 
\vid{\mu}{\displaystyle{\f{k}{2}}}\f{X_0(k)}{\omega(k)}
\rc\nonumber\\
&&\times \lc\f{\omega(k)+ X_{0}^{\dagger}(k)}{\omega(k)+b(k)}
\rc \lc\vi{\nu}{\displaystyle{\f{k}{2}}} +
\f{X_0(k)}{\omega(k)}\vid{\nu}{\displaystyle{\f{k}{2}}}\rc,\nonumber\\
\sigma_{\mu\nu}^{(b)}(k) &=& -\lc V^\dagger_{1\mu}\l(
  \f{k}{2}\r)V_{1\nu}\l(\f{k}{2}\r) + \f{1}{M_{0}}V_{1\mu}\l(
  \f{k}{2}\r)X_0^\dagger(k) V_{1\nu}\l(\f{k}{2}\r)\r.\nonumber \\
 &&  \hspace{0.5cm}+  V_{1\mu}\l(\f{k}{2}\r) V_{1\nu}^\dagger\l(\f{k}{2}\r)
\nonumber \\
&&  \hspace{0.5cm}-
  \l.\l(\f{1}{M_0} - \f{2}{\omega(k)}\r)V_{1\mu}^\dagger\l(\f{k}{2}\r)
X_0(k)V_{1\nu}^\dagger\l(\f{k}{2}\r)
\rc.\nonumber\\ \label{eq4} 
\end{eqnarray}
Now by the expressions of 
$X_0$, $V_{1\mu}$ and $\omega$
and the 
fact that there
are not doublers, singularity may occur around the region $k\simeq 0$
in each integral. Only the quark and gluon propagators can exhibit
singular behaviour for $k\simeq 0$. The leading order part in $k$
in $\Sigma^{(a)}(0)$ vanishes as  
\begin{eqnarray}
&& \f{\bar{g}^2}{aM_0}\sum_{\mu\nu}\int_{k^2 < \delta^2}\f{(\alpha -3)
{\ooalign{\hfil
/\hfil\crcr$k$}}}{k^4}=0.\quad\delta\ll 1.
\end{eqnarray}
Higher order terms in $k$ are obviously non-singular.
In $\Sigma^{(b)}(0)$ the part which can behave singular around
$k\simeq0$ is only the part of a gluon propagator and then 
the four momentum integral is non-singular. Since both $\Sigma^{(a)}(0)
$ and $\Sigma^{(b)}(0)$ is non-singular,
$\Sigma(p)$ does not have the divergences
of the forms $i\sla{p}C_{1n}/(a^2p^2)^n$ and
$ap^2C_{2n}/(a^2p^2)^{n+1}$. 
Now using
Eq.(\ref{eq1}) we can show that there is not the $1/a$ divergence 
in $\Sigma(p)$, and thus the $1/a$ divergence in $\Sigma^{a}$ and 
$\Sigma^{b}$ should cancel. Since the
highest order divergence is the $1/a$ divergence and 
derivatives 
with respect to momentum $p$ decrease the degree of
divergence, the $1/a$ divergent parts are given by Eq.(\ref{eq3}) and 
Eq.(\ref{eq4}).
We evaluate $\sigma_{\mu\nu}^{(a)}(k)$ and
$\sigma_{\mu\nu}^{(b)}(k)$, noting that odd functions of $k$
in them vanish. First, we investigate $\sigma_{\mu\nu}^{(b)}(k)$.
Because $\vid{\mu}{-k} = - \vi{\mu}{k}$ and $X_0^\dagger(-k) = X_0(k)$, 
we find $V_{1\mu}\l(\f{k}{2}\r)X_0^\dagger(k) 
V_{1\nu}\l(\f{k}{2}\r) - V_{1\mu}^\dagger\l(\f{k}{2}\r)
X_0(k)V_{1\nu}^\dagger\l(\f{k}{2}\r)$ is  odd .
Therefore
\begin{eqnarray}
\sigma_{\mu\nu}^{(b)}(k) &=& -\l( V^\dagger_{1\mu}V_{1\nu} +  
V_{1\mu} V_{1\nu}^\dagger + \f{2}{\omega}V_{1\mu}^\dagger
X_0 V_{1\nu}^\dagger\r).
\end{eqnarray}
We next evaluate $\sigma_{\mu\nu}^{(a)}(k)$. Using the relations 
$X_0X_0^\dagger = \omega^2$ and $X_0^2 = - \omega^2 + 2bX_0$, we obtain
\begin{eqnarray}
\sigma_{\mu\nu}^{(a)}(k) &=& V_{1\mu}V_{1\nu}^\dagger + V_{1\mu}^\dagger 
V_{1\nu}+ \f{1}{\omega + b}\lc V_{1\mu}X_0^\dagger V_{1\nu}
+ V_{1\mu}^\dagger X_0 V_{1\nu}^\dagger \r.\nonumber\\
&& + \omega \l(V_{1\mu}V_{1\nu} - V_{1\mu}^\dagger
V_{1\nu}^\dagger \r) + \f{2b}{\omega}V_{1\mu}^\dagger X_0 V_{1\nu}^\dagger 
+ V_{1\mu}i\gamma_{\rho}\sin{k_{\rho}}V_{1\nu}^\dagger \nonumber\\
&&\l.+ V_{1\mu}^\dagger i\gamma_{\rho}\sin{k_{\rho}}V_{1\nu}\rc.
\end{eqnarray}
Now $V_{1\mu}i\gamma_{\rho}\sin{k_{\rho}}V_{1\nu}^\dagger + V_{1\mu}^\dagger 
i\gamma_{\rho}\sin{k_{\rho}}V_{1\nu}$, $ V_{1\mu}V_{1\nu} - V_{1\mu}^\dagger V_{1\nu}^\dagger$ and
$V_{1\mu}X_0^\dagger V_{1\nu}+ V_{1\mu}^\dagger X_0 V_{1\nu}^\dagger$ are odd.
Therefore
\begin{eqnarray}
\sigma_{\mu\nu}^{(a)}(k) &=& V^\dagger_{1\mu}V_{1\nu} +  V_{1\mu} V_{1\nu}^\dagger+ \f{2}{\omega}V_{1\mu}^\dagger
X_0 V_{1\nu}^\dagger.
\end{eqnarray}
Thus $\Sigma^{(a)}(0)$ and $\Sigma^{(b)}(0)$ exactly cancel and
there is no $1/a$ divergence in $\Sigma(p)$. This analysis shows that
the massless pole of the Dirac operator at tree level is stable
against one-loop radiative correction\cite{CWZ}
\footnote{This part is studied previously in Ref. \cite{yamada}
and in Ref. \cite{domainwall}.}.

Next, we investigate the logarithmic divergence. 
From the above analysis $\Sigma(p)$ does not have the divergences
of the power of $a$. Therefore if there is the logarithmic divergence
in $\Sigma(p)$, it appears from the singular part in the integral 
for $a\to 0$ and the singular part is the integration region around
$k\simeq0$. $\Sigma^{(b)}(p)$ from which was subtracted the $1/a$
divergence
is non-singular for $k\simeq0$ and $a\to
0$ as it has only a gluon propagator. Accordingly we evaluate only
$\Sigma^{(a)}(p)$.
There are some ways for taking
out the logarithmic divergence~\cite{weak-coupling-exp,KS}. Here we use
the procedure discussed in the paper by Karsten and Smit. First, 
the denominators of the propagator 
are combined 
using Feynman's parameter as follows,

\begin{eqnarray} 
\Sigma^{(a)}(p) &=&   \f{\bar{g}^2}{2a}\sum_{\mu\nu}\int^1_0 dx\int_{k}
\lb\f{-2M_0\delta_
{\mu\nu}}{\lc -2M_0(1-x)(\omega(k)+b(k)) + x\widetilde{(pa-k)^2}\rc^2} 
\r.\nonumber\\
&& \l.- (1-\alpha)
\f{-4M_0x\widetilde{ (pa-k)}_\mu
  \widetilde{ (pa-k)_\nu}}{\lc -2M_0(1-x)(\omega(k)+b(k)) + 
x\widetilde{ (pa-k)^2}\rc^3}\rb\nonumber\\
&&\times\f{1}{\lc\omega(k)
  +\omega(pa)\rc^2}\lc\vi{\mu}{\displaystyle{\f{pa+k}{2}}} - 
\f{X_0(pa)}{\omega(pa)}\vid{\mu}{\displaystyle{\f{pa+k}{2}}}\f{X_0(k)}
{\omega(k)}
\rc\nonumber\\
&&\times \lc\vi{\nu}{\displaystyle{\f{pa+k}{2}}} -
\f{X_0(k)}{\omega(k)}\vid{\nu}{\displaystyle{\f{pa+k}{2}}}\f{X_0(pa)}
{\omega(pa)}\rc.\label{eq:x}
\end{eqnarray}
And the integration variables are shifted $k_\mu \to k_\mu +
axp_\mu$. Then we split the integral into two regions as follows,
\begin{eqnarray}
\int_{k^2} = \int_{k^2 <  \delta^2} + \quad\int_{k^2 > \delta^2} \qquad 
\delta\ll 1,
\end{eqnarray}
and we evaluate the $k^2 < \delta^2$ part in the continuum limit,
ignoring the $k^2 > \delta^2$ part which does not have the infrared 
divergence.
Eq.(\ref{eq:x}) is the complicated integral over the sines and cosines.
However for $k^2 < \delta^2$ and $a\to 0$ we can expand both 
the denominator and the numerator of Eq.(\ref{eq:x}) in $k$ and $a$.
The singular part corresponding to the logarithmic divergence
is obtained from the leading 
part in $k$ and $a$. 
Noting the spherical symmetry of the integral, the singular part is given by 
\begin{eqnarray}
&&\f{2i\bar{g}^2{\ooalign{\hfil/\hfil\crcr$p$}}}{M_0}\lb\int_0^1 dx x 
\int_{k^2 < \delta^2}\f{1}{\lc k^2 + p^2a^2x(1-x)\rc^2}\r.\nonumber\\ 
&&\l. + (1-\alpha)
\int_0^1 dx \l(-2x + \f{3}{2}x^2\r)\int_{k^2 < \delta^2}
\f{k^2}{\lc k^2 + p^2a^2x(1-x)\rc^3}\rb. \nonumber  
\end{eqnarray}
After some caluculations, the part proportional to
$\log{a^2p^2}$ is 
\begin{equation}
-\f{i\bar{g}^2{\ooalign{\hfil/\hfil\crcr$p$}}}{16\pi^2M_0}\lc1-(1-\alpha)\rc
\log{a^2p^2}.
\end{equation}
Now we pay attention to the normalization of quark fields.
 Since the continuum limit of $D_0$ is $i{\ooalign{\hfil/\hfil\crcr$p$}}/
|M_0|$, the effective action is 

\begin{equation}
S_{eff} = \int_p \bar{\psi}(p)\lc \f{i{\ooalign{\hfil/\hfil\crcr$p$}}}{|M_0|}
\l(1 - \f{\bar{g}^2}{16\pi^2}\alpha \log a^2p^2 + const\r)\rc\psi(p) + 
\cdots . 
\end{equation}
The coefficient $1/|M_0|$ can be absorbed by the redefinition 
of quark
fields. Finally, the divergent part of the quark
field renormalization factor $Z_\psi$ is 

\begin{equation}
Z_\psi = 1+\f{\bar{g}^2}{16\pi^2}
\alpha\log{a^2\mu^2},
\end{equation}
where $\mu$  is the renormalization scale. This factor agrees with
the continuum theory.

\section{Evaluation of the vacuum polarization }
\label{sec:gluon-self-energy}
\reseteqnum
In this section we calculate the one-loop vacuum polarization with
quark loop $\Pi_{\mu\nu}$ and then we show that there are not the
divergences of the forms  $p^2/(a^2p^2)^{n}\delta_{\mu\nu}$ 
and $p_{\mu}p_{\nu}/(a^2p^2)^{n}$ $(n\ge 1)$ 
and that the divergent part of the gluon field renormalization factor
is equal to that in the continuum theory.

Feynman diagrams for the vacuum polarization with quark
loop are shown in Fig.4.
\begin{figure}[htdp]
\begin{center}
\leavevmode
\epsfxsize=50mm
\epsfbox{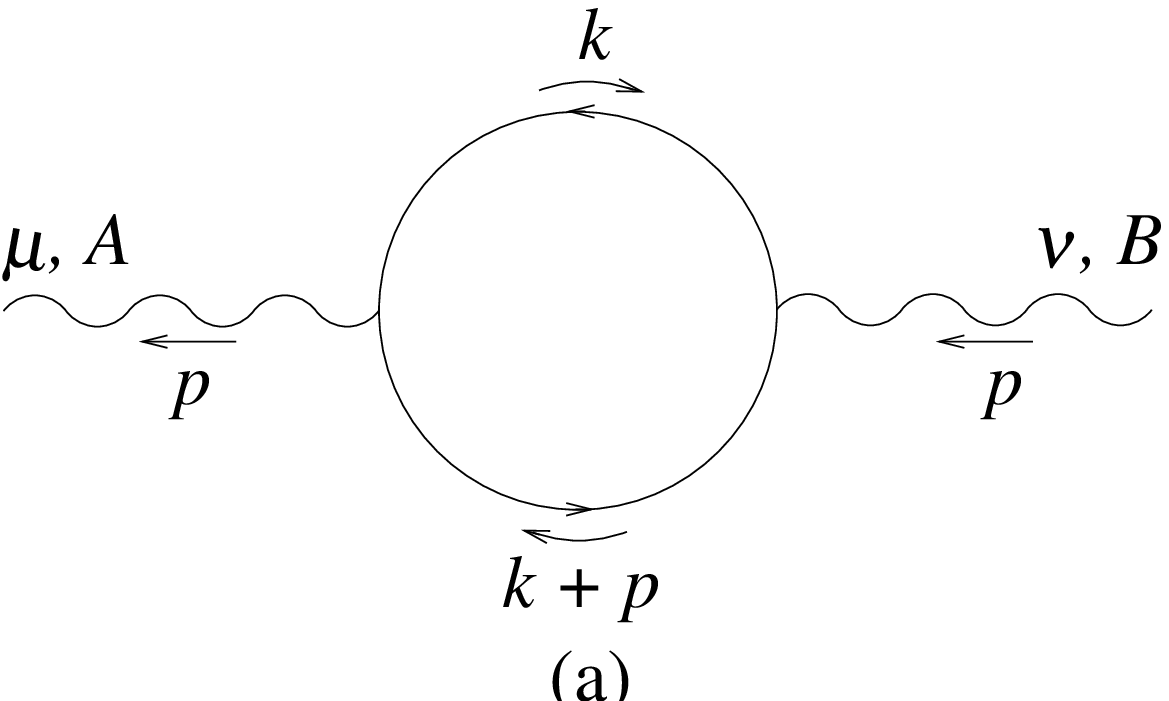}
\epsfxsize=50mm
\hspace{1.5cm}
\epsfbox{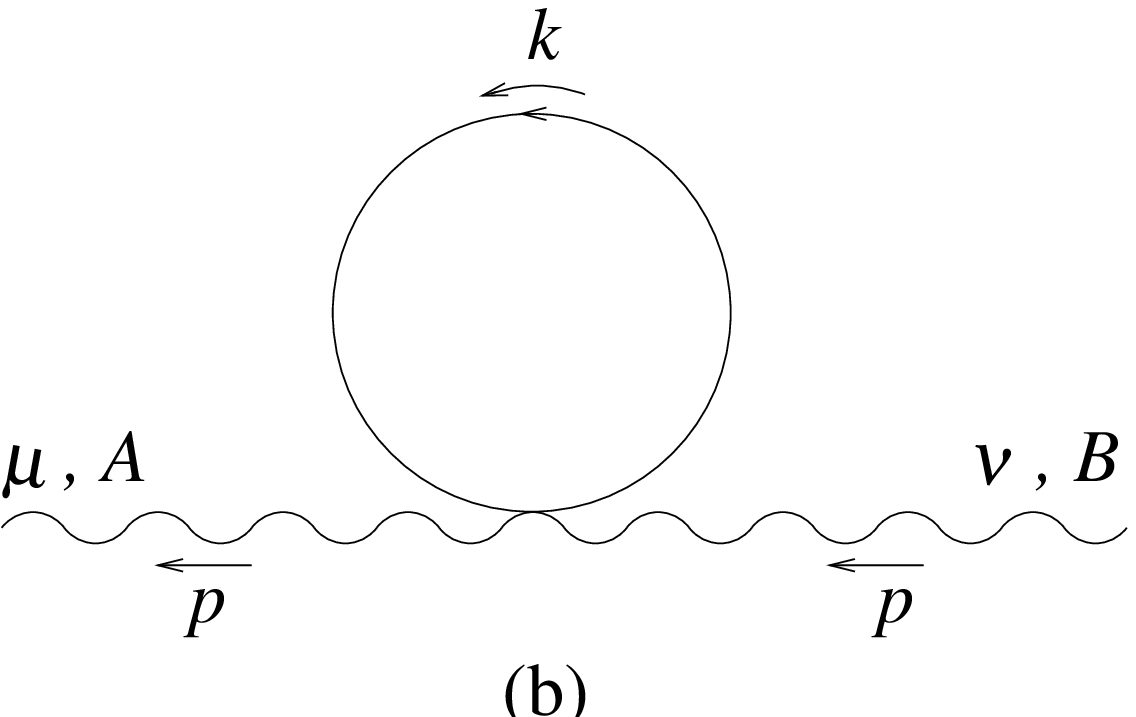}
\caption{Feynman diagrams for the vacuum polarization}
\label{fig:photon-se}
\end{center}
\end{figure}

The amplitude corresponding to Fig.4(a) is

\begin{eqnarray} 
\Pi_{\mu\nu}^{(a)}(p) &=&  - \f{g^2}{a^2}{\rm tr}\l(T^AT^B\r)
\intk\f{1}{\lc\omega(k)+\omega(k+p)\rc^2}
{\rm tr}\lb\lc\f{-i\gamma_{\rho}\sin{a(k+p)_{\rho}}}
{2(\omega(k+p)+b(k+p))}+\f{a}{2}\rc\r.\nonumber\\
&&\times\lc\vi{\nu}{\displaystyle{k+\f{p}{2}}} -
\f{X_0(k+p)}{\omega(k+p)}\vid{\nu}{\displaystyle{k+\f{p}{2}}}\f{X_0(k)}
{\omega(k)}\rc\lc\f{-i\gamma_{\rho}\sin{ak_{\rho}}}{2(\omega(k)+b(k))}
+\f{a}{2}\rc\nonumber\\
&&\times\l.\lc\vi{\mu}{\displaystyle{k+\f{p}{2}}} - 
\f{X_0(k)}{\omega(k)}\vid{\mu}{\displaystyle{k+\f{p}{2}}}
\f{X_0(k+p)}{\omega(k+p)}\rc\rb.\nonumber\\
\end{eqnarray}
The amplitude corresponding to Fig.4(b) is
\begin{eqnarray}
\Pi_{\mu\nu}^{(b)}(p) &=&  \f{g^2}{a^2}{\rm tr}\l(T^AT^B\r)
\intk {\rm tr}\lb\f{-i\gamma_{\rho}\sin{ak_{\rho}}}
{2(\omega(k)+b(k))}+\f{a}{2}\rb\nonumber\\
&&\times\lb\f{1}{2\omega(k)}\delta_{\mu\nu}\lc
\vii{\mu}{k} - \f{X_0(k)}{\omega(k)}
\viid{\mu}{k}\f{X_0(k)}{\omega(k)}\rc\r.\nonumber\\
&& - \f{1}{2\omega(k)(\omega(k) + \omega(k+p))^2}\lc
 X_0(k)\vid{\mu}{\kph}\vi{\nu}{\kph}\r.\nonumber\\
&& + \vi{\mu}{\kph}X_0^\dagger(k+p)\vi{\nu}{\kph}
+ \vi{\mu}{\kph}\vid{\nu}{\kph}X_0(k)\nonumber\\
&& \l.- \f{2\omega(k) + \omega(k+p)}{\omega(k)^2
  \omega(k+p)}X_0(k)\vid{\mu}{\kph}X_0(k+p)\vid{\nu}{\kph}X_0(k)\rc\nonumber\\
&& - \f{1}{2\omega(k)(\omega(k) + \omega(k-p))^2}\lc
 X_0(k)\vid{\nu}{\kphm}\vi{\mu}{\kphm}\r.\nonumber\\
&& + \vi{\nu}{\kphm}X_0^\dagger(k-p)\vi{\mu}{\kphm}
+ \vi{\nu}{\kphm}\vid{\mu}{\kphm}X_0(k)\nonumber\\
&& \l.\l.- \f{2\omega(k) + \omega(k-p)}{\omega(k)^2
  \omega(k-p)}X_0(k)\vid{\nu}{\kphm}X_0(k-p)\vid{\mu}{\kphm}X_0(k)\rc\rb.
\nonumber\\
\end{eqnarray}
Rescaling the integration momentums as before $k_{\mu}\to k_{\mu}/a$ and noting
that for $N_f$ flavor QCD, $tr\l(T^AT^B\r) = N_f\delta^{AB}$, 
we obtain
\begin{eqnarray} 
\Pi_{\mu\nu}^{(a)}(p) 
&=&  - \f{N_fg^2}{4a^2}\int_{k}\f{1}{\lc\omega(k)+\omega(k+pa)\rc^2}
{\rm tr}\lb\lc\f{-i\gamma_{\rho}\sin{(k+pa)_{\rho}}}
{(\omega(k+pa)+b(k+pa))}+ 1\rc\r.\nonumber\\
&&\times\lc\vi{\nu}{\displaystyle{k+\f{pa}{2}}} -
\f{X_0(k+pa)}{\omega(k+pa)}\vid{\nu}{\displaystyle{k+\f{pa}{2}}}\f{X_0(k)}
{\omega(k)}\rc\lc\f{-i\gamma_{\rho}\sin{k_{\rho}}}{(\omega(k)+b(k))}
+ 1\rc\nonumber\\
&&\times\l.\lc\vi{\mu}{\displaystyle{k+\f{pa}{2}}} - 
\f{X_0(k)}{\omega(k)}\vid{\mu}{\displaystyle{k+\f{pa}{2}}}
\f{X_0(k+pa)}{\omega(k+pa)}\rc\rb,\label{eq:Pi^a}
\end{eqnarray}
and 

\begin{eqnarray}
\Pi_{\mu\nu}^{(b)}(p) &=&  \f{N_fg^2}{4a^2}\int_k \f{1}{\omega(k)}
{\rm tr}\lb\f{-i\gamma_{\rho}\sin{k_{\rho}}}
{(\omega(k)+b(k))}+ 1 \rb\nonumber\\
&&\times\lb\delta_{\mu\nu}\lc
\vii{\mu}{k} - \f{X_0(k)}{\omega(k)}
\viid{\mu}{k}\f{X_0(k)}{\omega(k)}\rc\r.\nonumber\\
&& - \f{1}{(\omega(k) + \omega(k+pa))^2}\lc
 X_0(k)\vid{\mu}{\displaystyle{k+\f{pa}{2}}}\vi{\nu}
{\displaystyle{k+\f{pa}{2}}}\r.\nonumber\\
&& + \vi{\mu}{\displaystyle{k+\f{pa}{2}}}X_0^\dagger(k+pa)\vi{\nu}
{\displaystyle{k+\f{pa}{2}}}+ \vi{\mu}{\displaystyle{k+\f{pa}{2}}}
\vid{\nu}{\displaystyle{k+\f{pa}{2}}}X_0(k)\nonumber\\
&& \l.- \f{2\omega(k) + \omega(k+pa)}{\omega(k)^2
  \omega(k+pa)}X_0(k)\vid{\mu}{\displaystyle{k+\f{pa}{2}}}X_0(k+pa)
\vid{\nu}{\displaystyle{k+\f{pa}{2}}}X_0(k)\rc\nonumber\\
&& - \f{1}{(\omega(k) + \omega(k-pa))^2}\lc
 X_0(k)\vid{\nu}{\displaystyle{k-\f{pa}{2}}}\vi{\mu}
{\displaystyle{k-\f{pa}{2}}}\r.\nonumber\\
&& + \vi{\nu}{\displaystyle{k-\f{pa}{2}}}X_0^\dagger(k-pa)\vi{\mu}
{\displaystyle{k-\f{pa}{2}}}+ \vi{\nu}{\displaystyle{k-\f{pa}{2}}}\vid{\mu}
{\displaystyle{k-\f{pa}{2}}}X_0(k)\nonumber\\
&& \l.\l.- \f{2\omega(k) + \omega(k-pa)}{\omega(k)^2
  \omega(k-pa)}X_0(k)\vid{\nu}{\displaystyle{k-\f{pa}{2}}}X_0(k-pa)
\vid{\mu}{\displaystyle{k-\f{pa}{2}}}X_0(k)\rc\rb,\nonumber\\
\end{eqnarray}
where $\omega$, $X_0$, $V_{1\mu}$,
$V_{2\mu}$ and $G_{\mu\nu}$ are appropriately redefined as done in the
previous section.
First, we show that
$\Pi_{\mu\nu}^{(a)}(p)$ and 
$\Pi_{\mu\nu}^{(b)}(p)$ are constants for $p=0$ and thus the divergent
terms of the forms $p^2/(a^2p^2)^{n+1}\delta_{\mu\nu}$ and 
$p_{\mu}p_{\nu}/(a^2p^2)^{n}$ do not appear.
 Setting $p=0$ in
them, we have
\begin{eqnarray} 
\label{eq20}
\Pi_{\mu\nu}^{(a)}(0) &=&  - \f{N_fg^2}{16a^2}\int_{k}\f{1}{\omega(k)^2}
{\rm tr}\lb\lc\f{-i\gamma_{\rho}\sin{ak_{\rho}}}
{(\omega(k)+b(k))}+ 1\rc\r.\nonumber\\
&&\times\lc\vi{\nu}{\displaystyle{k}} -
\f{X_0(k)}{\omega(k)}\vid{\nu}{\displaystyle{k}}\f{X_0(k)}
{\omega(k)}\rc\lc\f{-i\gamma_{\rho}\sin{k_{\rho}}}{(\omega(k)+b(k))}
+ 1\rc\nonumber\\
&&\times\l.\lc\vi{\mu}{\displaystyle{k}} - 
\f{X_0(k)}{\omega(k)}\vid{\mu}{\displaystyle{k}}
\f{X_0(k)}{\omega(k)}\rc\rb,\nonumber\\
\end{eqnarray}   
and 
\begin{eqnarray}
\label{eq21}
\Pi_{\mu\nu}^{(b)}(0) &=&  \f{N_fg^2}{4a^2}\int_k \f{1}{\omega(k)}
{\rm tr}\lb\f{-i\gamma_{\rho}\sin{k_{\rho}}}
{(\omega(k)+b(k))}+ 1 \rb\nonumber\\
&&\times\lb\delta_{\mu\nu}\lc
\vii{\mu}{k} - \f{X_0(k)}{\omega(k)}
\viid{\mu}{k}\f{X_0(k)}{\omega(k)}\rc\r.\nonumber\\
&& - \f{1}{4\omega(k)^2}\lc
 X_0(k)\vid{\mu}{\displaystyle{k}}\vi{\nu}
{\displaystyle{k}}\r.\nonumber\\
&& + \vi{\mu}{\displaystyle{k}}X_0^\dagger(k)\vi{\nu}
{\displaystyle{k}}+ \vi{\mu}{\displaystyle{k}}
\vid{\nu}{\displaystyle{k}}X_0(k)\nonumber\\
&& \l.- \f{3}{\omega(k)^2}X_0(k)\vid{\mu}{\displaystyle{k}}X_0(k)
\vid{\nu}{\displaystyle{k}}X_0(k)\rc\nonumber\\
&& - \f{1}{4\omega^2}\lc
 X_0(k)\vid{\nu}{\displaystyle{k}}\vi{\mu}
{\displaystyle{k}}\r.\nonumber\\
&& + \vi{\nu}{\displaystyle{k}}X_0^\dagger(k)\vi{\mu}
{\displaystyle{k}}+ \vi{\nu}{\displaystyle{k}}\vid{\mu}
{\displaystyle{k}}X_0(k)\nonumber\\
&& \l.\l.- \f{3}{\omega(k)^2}X_0(k)\vid{\nu}{\displaystyle{k}}X_0(k)
\vid{\mu}{\displaystyle{k}}X_0(k)\rc\rb.
\end{eqnarray}
In $\Pi_{\mu\nu}^{(b)}(0)$ the part which can be singular is only 
a quark 
propagator which behaves $i\sla{k}/k^2$ at $k\simeq 0$ and the four 
momentum integral is non-singular, and thus the results of the $k$ 
integration is a constant. In 
$\Pi_{\mu\nu}^{(a)}(0)$ the leading part in $k$ is

\begin{eqnarray}
&&  \f{N_fg^2}{4a^2}\int_{k^2 < \delta^2} 
\f{{\rm tr}\lb{\ooalign{\hfil/\hfil
\crcr$k$}}(2i\gamma_\nu){\ooalign{\hfil/\hfil\crcr$k$}}(2i\gamma_\mu)\rb}
{k^4}\nonumber\\ 
&& \sim \f{1}{a^2}\int_0^\delta dk k .\nonumber
\end{eqnarray}
Hence $\Pi_{\mu\nu}^{(a)}(0)$ is a constant as well. The above 
analysis exhibits 
that there is not
the divergences of the forms $p^2/(a^2p^2)^{n+1}\delta_{\mu\nu}$ and 
$p_{\mu}p_{\nu}/(a^2p^2)^{n}$ 
in $\Pi_{\mu\nu}(p)$. Accordingly assuming that $SO(4)$ symmetry
restores in the continuum limit, the structure of the vacuum
polarization amplitude is

\begin{eqnarray}
&& \Pi_{\mu\nu}(p) = \f{1}{a^2}C_0 \delta_{\mu\nu} + C_{1\mu\nu}(p) + O(a^2).
   \qquad C_0 : constant.
\end{eqnarray}
However we take the gauge invariant formulation and the Ward
identity determines the structure of $\Pi_{\mu\nu}(p)$ as follows,
  
\begin{eqnarray}
&& \Pi_{\mu\nu}(p) = (p^2\delta_{\mu\nu} - p_\mu p_\nu)\Pi(p^2).
\end{eqnarray}
From the above two equations we conclude that there is not also the
$1/a^2$ divergence in $\Pi_{\mu\nu}(p)$.

Next, we evaluate the logarithmic divergent part in the vacuum
polarization. 
The procedure which taking out the logarithmic divergent part is same 
as we used in the previous section. Since $\Pi_{\mu\nu}^{(b)}(p)$ 
from which was subtracted the $1/a^2$ divergence
is not singular for $k\simeq 0$ and $a\to 0$ from the same reason for 
$\Sigma^{(b)}(p)$, 
we consider only
$\Pi_{\mu\nu}^{(a)}(p)$. In Eq.(\ref{eq:Pi^a}), 
using Feynman's parameter and shifting 
the integration variables $k_\mu \to k_\mu - axp_\mu $ and 
then we evaluate the part of the integration region 
$k^2 < \delta^2, \;\delta \ll
1$ in the continuum limit. The singular part for $k\simeq 0$ and 
$a\to0$ is the leading part in $k$ and $a$.
Noting the spherical symmetry of the integral,
its part is given by
\begin{eqnarray} 
&&  \f{N_fg^2}{a^2}\int_0^1 dx\int_{k^2 < \delta^2}
\f{2k^2\delta_{\mu\nu} - 4a^2x(1-x)(p^2\delta_{\mu\nu}-2p_\mu p_\nu)}
{\l(k^2 + p^2a^2x(1-x)\r)^2}.
\end{eqnarray}
After some calculations, the term proportional to 
$\log a^2p^2$ is extracted as follows,
\begin{eqnarray}
&& \f{N_fg^2}{12\pi^2}(p^2\delta_{\mu\nu}- p_\mu p_\nu)\log a^2p^2.
\end{eqnarray}
Hence the divergent part of the gluon field renormalization factor 
due to the quark loop
is written as 
$Z_{A} = 1+(N_fg^2/12\pi^2)
\log a^2\mu^2$, which  agrees with the continuum theory.

\section{Summary and discussion}
\label{sec:discussion}

In this paper, we discussed the weak coupling expansion of  
massless QCD defined with the overlap Dirac operator at one-loop. 
In the weak coupling expansion of the overlap Dirac operator,
the fermion propagator has the single pole for the physical mode
and is free from species doublers.
The quark-gluon vertices are regular functions in momenta
and thus the gauge interaction is local.
With these properties of the quark propagator and 
the quark-gluon verteces, we have studied the fermion self-energy 
and vacuum polarization and confirmed that non-local divergent terms 
are not in fact induced, 
which are not necessarily forbidden by the Ginsparg-Wilson relation.
Then the Ginsparg-Wilson relation prohibits 
linearly divergent mass terms.
The divergent parts of the wave function renormalization factors 
for quarks and gluons agree with the continuum theory.

\end{document}